\documentclass[12pt,article,onecolumn]{IEEEtran}

\usepackage{srcltx,epsf,amsmath,amssymb,amsfonts,mathrsfs}
\usepackage{times}
\usepackage{subeqnarray}
\usepackage{epsfig}
\usepackage{url}

\newtheorem{theorem}{Theorem}

\def \md {{\rm\:mod\:} \Lambda}

\begin{document}

% paper title
%\title{Capacity Bounds for the Gaussian Two-way Relay Channel}
\title{Capacity of the Gaussian Two-way Relay Channel to
within $\frac{1}{2}$ Bit}
%On the Capacity of the Gaussian Two-way Relay Channel to within
%$\frac{1}{2}$ Bit}
\author{\authorblockN{Wooseok Nam, \IEEEmembership{Student Member, IEEE,}
        Sae-Young Chung, \IEEEmembership{Senior Member, IEEE,}
        and Yong H. Lee, \IEEEmembership{Senior Member, IEEE}}\\
\authorblockA{School of EECS, KAIST,\\ Daejeon, Republic of Korea\\
E-mail: wsnam@stein.kaist.ac.kr, sychung@ee.kaist.ac.kr,
yohlee@ee.kaist.ac.kr}\thanks{This work was supported by the IT R\&D
program of MKE/IITA. [2008-F-004-01, 5G mobile communication systems
based on beam division multiple access and relays with group
cooperation]} } \maketitle

\begin{abstract}
In this paper, a Gaussian two-way relay channel, where two source
nodes exchange messages with each other through a relay, is
considered. We assume that all nodes operate in full-duplex mode and
there is no direct channel between the source nodes. We propose an
achievable scheme composed of nested lattice codes for
the uplink and structured binning for the downlink. %Using structured
%codes for the uplink provides the gain of computation coding and is
%closely related to the concept of network coding for wireless
%networks.
We show that the scheme achieves within $\frac{1}{2}$ bit from the
cut-set bound for all channel parameters and becomes asymptotically
optimal as the signal to noise ratios increase.
\end{abstract}


\section{Introduction} \label{SEC:Introduction}

\begin{figure} [t]
\begin{center}
\epsfig{file=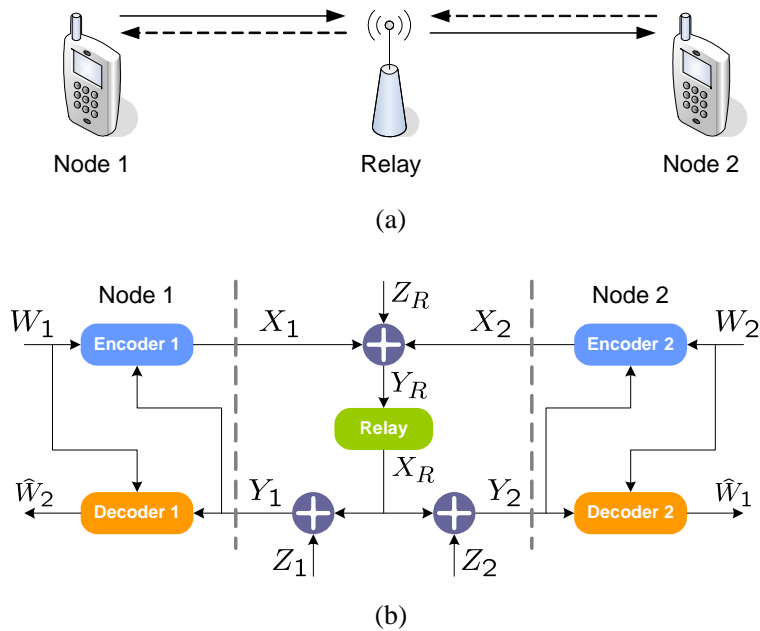, width=10cm} \caption{Gaussian two-way
relay channel} \label{FIG:TRC_gau}
\end{center}
\end{figure}

We consider a two-way relay channel (TRC), as shown in Fig.
\ref{FIG:TRC_gau} (a). Nodes 1 and 2 want to exchange messages with
each other, and a relay node facilitates the communication between
them. This TRC can be thought of as a basic building block of
general wireless networks, along with the relay channel
\cite{CoverIT79}, the two-way channel \cite{ShannonTWC61}, etc.
Recently, there have been a great deal of interest in the capacity
of wireless networks. Inspired by network coding \cite{Ahlswede},
TRC has been studied in the context of network coding for wireless
networks due to its simple structure. However, the capacity region
of the general TRC is still unknown.

In \cite{RankovISIT06}, several classical relaying strategies for
the one-way relay channel \cite{CoverIT79}, such as
amplify-and-forward (AF), decode-and-forward (DF), and
compress-and-forward (CF), were extended and applied to the TRC. AF
relaying is a very simple and practical strategy, but due to the
noise amplification, it cannot be optimal in throughput at low
signal to noise ratios (SNRs). DF relaying %see the TRC as a
%combination of multiple access channel (MAC) and broadcast channel
%(BC), and
requires the relay to decode all the source messages and, thus, does
not suffer from the noise amplification. In \cite{KnoppIZS06}, it
was shown that the achievable rate region of DF relaying can be
improved by applying network coding to the decoded messages at the
relay. This scheme is sometimes optimal in its throughput
\cite{OechteringIT08}, but it is generally subject to the {\em
multiplexing loss} \cite{KnoppISIS07}.

In general, in relay networks, the relay nodes need not reconstruct
all the messages, but only need to pass sufficient information to
the destination nodes to do so. CF or partial DF relaying strategies
for the TRC, in which the relay does not decode the source messages,
were studied in \cite{AvestimehrAllerton08, GunduzAllerton08}. It
was shown that these strategies achieve the information theoretic
cut-set bound \cite{CoverText} within a constant number of bits when
applied to the Gaussian TRC. In \cite{NamIZS08, Narayanan},
structured schemes that use lattice codes were proposed for the
Gaussian TRC, and it was shown that these schemes can achieve the
cut-set bound within $\frac{1}{2}$ bit.

In this paper, we focus on the Gaussian TRC with full-duplex nodes
and no direct communication link between the source nodes. Such a
Gaussian TRC is shown in Fig. \ref{FIG:TRC_gau} (b), and it is
essentially the same as those considered in
\cite{AvestimehrAllerton08}-\cite{Narayanan}. For the uplink, i.e.,
the channel from the source nodes to the relay, we propose a scheme
based on nested lattice codes \cite{ErezIT04} formed from a lattice
chain. This scheme is borrowed from the work on the relay networks
with interference in \cite{NamAllerton08, NamIT09Net}. By using
nested lattice codes for the uplink, we can exploit the structural
gain of {\em computation coding} \cite{NazerIT07}, which corresponds
to a kind of combined channel and network coding. For the downlink,
i.e., the channel from the relay to the destination nodes, we see
the channel as a BC with receiver side information
\cite{OechteringIT08, XieCWIT07, WuISIT07}, since the receiver nodes
know their own transmitted messages. In such a channel, the capacity
region can be achieved by the random binning of messages
\cite{XieCWIT07}. In our strategy, a structural binning of messages,
rather than the random one, is naturally introduced by the lattice
codes used in the uplink. Thus, at each destination node, together
with the side information on its own message, this binning can be
exploited for decoding.

In fact, as stated above, our work is not the first to apply lattice
codes to the Gaussian TRC. However, we assume more a general TRC
model compared to the other works. In \cite{Narayanan}, it was
assumed that the channel is symmetric, i.e., all source and relay
nodes have the same transmit powers and noise variances. In
\cite{NamIZS08}, a lattice scheme for the asymmetric Gaussian TRC
was proposed. However, the scheme requires the existence of a
certain class of lattices to achieve a $\frac{1}{2}$ bit gap to the
cut-set bound. This paper extends those previous works and shows
that we can in fact achieve the cut-set bound within $\frac{1}{2}$
bit\footnote{To be exact, this implies $\frac{1}{2}$ bit per
dimension. For a complex-valued system, as considered in
\cite{AvestimehrAllerton08}, we have 1 bit gap per complex
dimension.} for any channel parameters, e.g., transmit powers and
noise variances. Moreover, the gap vanishes as the uplink SNRs
increase.

This paper is organized as follows. In Section \ref{SEC:Model}, we
present the channel model and define related parameters. The cut-set
bound on the capacity region is given in Section \ref{SEC:Cutset}.
Section \ref{SEC:ACH} illustrates our achievable scheme and computes
the achievable rate region. Section \ref{SEC:Conclusion} concludes
the paper.

\psfull
\psdraft

\section{System Model} \label{SEC:Model}

We consider a Gaussian two-way relay channel, as shown in Fig.
\ref{FIG:TRC_gau} (b). We assume that the source and relay nodes
operate in full-duplex mode and there is no direct path between the
two source nodes. The variables of the channel are as follows:
\begin{itemize}
\item $W_i \in \left\{1,\ldots,2^{nR_i} \right\}$: message
of node $i$,
\item ${\bf X}_i=\left[X_i^{(1)},\ldots,X_i^{(n)}\right]^T$: channel input of
node $i$,
\item ${\bf Y}_R=\left[Y_R^{(1)},\ldots,Y_R^{(n)}\right]^T$: channel output at the
relay,
\item ${\bf X}_R=\left[X_R^{(1)},\ldots,X_R^{(n)}\right]^T$:
channel input of the relay,
\item ${\bf Y}_i=\left[Y_i^{(1)},\ldots,Y_i^{(n)}\right]^T$: channel output at node
$i$,
\item $\hat{W}_i \in \left\{1,\ldots,2^{nR_i
} \right\}$: estimated message of node $i$,
\end{itemize}
where $i\in\{1,2\}$, $n$ is the number of channel uses, and $R_i$
denotes the rate of node $i$. We assume that the messages $W_1$ and
$W_2$ are independent of each other.
 %At time $t$, node $i$ encode its transmit signal
%as a function of message and past received signal, i.e.,
%$X_i^{(t)}=f_i^{(t)}\left(W_i,Y_i^{t-1}\right)$, where
%$Y_i^{t-1}=\left\{Y_i^{(1)},\ldots,Y_i^{(t-1)} \right\}$.
%
%According to the message and the past received signal, node $i$
%encode its transmit signal for time $t$
Node $i$ transmits $X_i^{(t)}$ at time $t$ to the relay through the
uplink channel specified by
\begin{equation*}
Y_R^{(t)} = X_1^{(t)} + X_2^{(t)} + Z_R^{(t)} \text,
\label{EQ:RXsig}
\end{equation*}
where $Z_R^{(t)}$ is an independent identically distributed (i.i.d.)
Gaussian random variable with zero mean and variance $\sigma_R^2$.
%The input alphabet of the channel is the field of real numbers, and
The transmit signal $X_i^{(t)}$ is determined as a function of
message $W_i$ and past channel outputs
$Y_i^{t-1}=\left\{Y_i^{(1)},\ldots,Y_i^{(t-1)} \right\}$, i.e.,
$X_i^{(t)}=f_i^{(t)}\left(W_i,Y_i^{t-1}\right)$. There are power
constraints $P_i$, $i\in\{1,2\}$ on the transmitted signals
\begin{equation*}
\frac{1}{n} \sum_{t=1}^n \left( X_i^{(t)} \right)^2 \leq P_i,\;
i=1,2 \text.
\end{equation*}
At the same time, the relay transmits $X_R^{(t)}$ to nodes 1 and 2
through the downlink channel specified by
\begin{equation*}
Y_i^{(t)} = X_R^{(t)} + Z_i^{(t)},\; i \in \{1,2\} \text,
\end{equation*}
where $Z_i^{(t)}$ is an i.i.d. Gaussian random variable with zero
mean and variance $\sigma_i^2$. The power constraint at the relay is
given by
\begin{equation*}
\frac{1}{n} \sum_{t=1}^n \left( X_R^{(t)} \right)^2 \leq P_R \text.
\end{equation*}
Since the relay has no messages of its own, $X_R^{(t)}$ is formed as
a function of past channel outputs $Y_R^{t-1}= \{
Y_R^{(1)},\ldots,Y_R^{(t-1)} \}$, i.e., $X_R^{(t)}=f_R^{(t)}\left(
Y_R^{t-1} \right)$. At node $1$, the message estimate $\hat{W}_2 =
g_1(W_1,{\bf Y}_1)$ is computed from the received signal ${\bf Y}_1$
and its message $W_1$. The decoding of node $2$ is performed
similarly. Now, the average probability of error is defined as
\begin{equation*}
P_e = \Pr \left\{ \hat{W}_1 \neq W_1 \text{ or } \hat{W}_2 \neq W_2
\right\} \text.
\end{equation*}
For the aforementioned TRC, we say that a rate pair $(R_1,R_2)$ is
achievable if a sequence of encoding and decoding functions exists
such that the error probability vanishes as $n$ tends to infinity.
The capacity region of the TRC is defined as the convex closure of
all achievable rate pairs.

\section{An upper bound for the capacity region} \label{SEC:Cutset}

By the cut-set bound \cite{CoverText},  if the rate pair $(R_1,
R_2)$ is achievable for a general TRC, a joint probability
distribution $p(x_1, x_2, x_R)$ exists such that
\begin{subeqnarray}
&R_1 \leq \min \left\{ I(X_1;Y_R,Y_2|X_R,X_2), I(X_1,X_R;Y_2|X_2)
\right\} \text, \\
&R_2 \leq \min \left\{ I(X_2;Y_R,Y_1|X_R,X_1), I(X_2,X_R;Y_1|X_1)
\right\} \text. \label{EQ:Cutset}
\end{subeqnarray}
In particular, for the Gaussian TRC, we can use the fact that there
is no direct path between nodes 1 and 2, i.e., $(X_1,X_2,Y_R)
\rightarrow X_R \rightarrow (Y_1, Y_2)$, and that $X_R \rightarrow
(X_1,X_2) \rightarrow Y_R$. This induces another upper bound from
(\ref{EQ:Cutset}), given by
\begin{subeqnarray}
&R_1 \leq \min \left\{ I(X_1;Y_R|X_2), I(X_R;Y_2)
\right\} \text, \\
&R_2 \leq \min \left\{ I(X_2;Y_R|X_1), I(X_R;Y_1) \right\} \text,
\label{EQ:CutsetSimp}
\end{subeqnarray}
for some $p(x_1,x_2,x_R)$. It can be easily seen that, for the
Gaussian TRC with transmit power constraints,
%If we consider the first term under the minimization in
%(\ref{EQ:Cutset}a),
%\begin{align}
%I(X_1;Y_R,Y_2|X_R,X_2) &= I(X_1;Y_R|X_R,X_2) \nonumber\\
%&\leq I(X_1;Y_R|X_2) \text,
%\end{align}
%where the first equality follows from the fact that there is no
%direct path between the source nodes, i.e., $(X_1,X_2,Y_R)
%\rightarrow X_R \rightarrow Y_2$, and the second inequality from
%$X_R \rightarrow (X_1,X_2) \rightarrow Y_R$. For the second term
%under the minimization in (\ref{EQ:Cutset}a), we have
%\begin{equation}
%I(X_1,X_R;Y_2|X_2) \leq I(X_R;Y_2) \text,
%\end{equation}
%where the inequality follows from the Markov chain, $(X_1,X_2)
%\rightarrow X_R \rightarrow Y_2$. We can do the same for
%(\ref{EQ:Cutset}b), and thus we have
%\begin{subeqnarray}
%&R_1 \leq \min \left\{ I(X_1;Y_R|X_2), I(X_R;Y_2)
%\right\} \text, \\
%&R_2 \leq \min \left\{ I(X_2;Y_R|X_1), I(X_R;Y_1) \right\} \text,
%\label{EQ:CutsetSimp}
%\end{subeqnarray}
%for some $p(x_1,x_2,x_R)$. In particular, for the Gaussian TRC,
all terms under the minimizations in (\ref{EQ:CutsetSimp}) are
maximized by the product distribution $p(x_1,x_2,x_R) =
p(x_1)p(x_2)p(x_R)$, where $p(x_1)$, $p(x_2)$, and $p(x_R)$ are
Gaussian probability density functions with zero means and variances
$P_1$, $P_2$, and $P_R$, respectively. The resulting upper bound on
the capacity region is given by
\begin{subequations}
\begin{align}
&R_1 \leq \min \left\{ \frac{1}{2} \log \left( 1+
\frac{P_1}{\sigma_R^2} \right), \frac{1}{2} \log \left( 1+
\frac{P_R}{\sigma_2^2} \right) \right\}\text, \\
&R_2 \leq \min \left\{ \frac{1}{2} \log \left( 1+
\frac{P_2}{\sigma_R^2} \right), \frac{1}{2} \log \left( 1+
\frac{P_R}{\sigma_1^2} \right) \right\}\text.
\end{align}
\label{EQ:CutsetGaussian}
\end{subequations}

\section{An achievable rate region for the Gaussian TRC}
\label{SEC:ACH}

In this section, we compute an achievable rate region for the
Gaussian TRC. For the uplink, we consider using nested lattice
codes, which are formed from a lattice chain. For the downlink, we
use a structured binning of messages at the relay, which is
naturally introduced by the nested lattice codes. The destination
nodes decode each other's message using this binning and the side
information on their own transmitted messages.

The main result of this section is as follows:

\begin{theorem}
For a Gaussian TRC, as shown in Fig. \ref{FIG:TRC_gau} (b), we can
achieve the following region:
\begin{subequations}
\begin{align}
R_1 &\leq \min \left\{  \left[ \frac{1}{2} \log \left(
\frac{P_1}{P_1 + P_2} + \frac{P_1}{\sigma_R^2} \right) \right]^+,
\frac{1}{2} \log
\left( 1+ \frac{P_R}{\sigma_2^2} \right) \right\} \text, \\
R_2 &\leq \min \left\{ \left[ \frac{1}{2} \log \left( \frac{P_2}{P_1
+ P_2} + \frac{P_2}{\sigma_R^2} \right) \right]^+, \frac{1}{2} \log
\left( 1+ \frac{P_R}{\sigma_1^2} \right) \right\} \text,
\end{align}
\label{EQ:AchRegion}
\end{subequations}
where $[x]^+ \triangleq \max \{x,0\}$.
 \label{TH:TRCACH}
\end{theorem}

Note that the achievable rate region in Theorem \ref{TH:TRCACH} is
within $\frac{1}{2}$ bit of the upper bound
(\ref{EQ:CutsetGaussian}), regardless of channel parameters such as
the transmit powers and noise variances. Moreover, as the uplink
SNRs $\frac{P_1}{\sigma_R^2}$ and $\frac{P_2}{\sigma_R^2}$ increase,
the gap vanishes and our achievable region asymptotically approaches
the capacity region of the Gaussian TRC.

We prove Theorem \ref{TH:TRCACH} in the following subsections.

\subsection{Lattice scheme for the uplink}

For the scheme for the uplink, we consider a lattice coding scheme.
We will not cover the full details of lattices and lattice codes due
to page limitations. For a comprehensive review, we refer the reader
to \cite{ErezIT04}-\cite{ForneyAllerton03} and the references
therein.

A {\em nested lattice code} is defined in terms of two
$n$-dimensional lattices $\Lambda_C^n$ and $\Lambda^n$, which form a
lattice partition $\Lambda_C^n / \Lambda^n$, i.e., $\Lambda^n
\subseteq \Lambda_C^n$. The nested lattice code is a lattice code
which uses $\Lambda_C^n$ as codewords and the Voronoi region of
$\Lambda^n$ as a shaping region. For $\Lambda_C^n / \Lambda^n$, we
define the {\em set of coset leaders} as
\begin{equation*}
\mathcal{C} = \{ \Lambda_C^n \md^n \} \triangleq \{ \Lambda_C^n \cap
\mathcal{R} \} \text,
\end{equation*}
where $\mathcal{R}$ is the Voronoi region of $\Lambda$. Then the
coding rate of the nested lattice code is given by
\begin{equation*}
R = \frac{1}{n} \log | \mathcal{C} | = \frac{1}{n} \log \frac{{\rm
Vol}(\Lambda^n)}{{\rm Vol}(\Lambda_C^n)} \text,
\end{equation*}
where ${\rm Vol}(\cdot)$ denotes the volume of the Voronoi region of
a lattice. For the TRC, we should design two nested lattice codes,
one for each source node. This subsection will show how the nested
lattice codes are formed. In the following argument, we assume that
$P_1 \geq P_2$ without loss of generality. Now, let us first
consider a theorem that is a key for our code construction.

\begin{theorem}
For any $P_1 \geq P_2 \geq 0$ and $\gamma \geq 0$, a sequence of
$n$-dimensional lattice chains $\Lambda_1^{n} \subseteq
\Lambda_2^{n}\subseteq \Lambda_C^{n}$ exists that satisfies the
following properties.

\noindent a) $\Lambda_1^{n}$ and $\Lambda_2^{n}$ are simultaneously
Rogers-good and Poltyrev-good while $\Lambda_C^{n}$ is Poltyrev-good
(for the notion of goodness of lattices, see \cite{ErezIT05}).

\noindent b) For any $\epsilon > 0$, $P_i - \epsilon \leq \sigma^2
(\Lambda_i^n) \leq P_i$, $i \in \{1,2\}$, for sufficiently large
$n$, where $\sigma^2(\cdot)$ denotes the second moment per dimension
associated with the Voronoi region of the lattice.

\noindent c) The coding rate of the nested lattice code associated
with the lattice partition $\Lambda_C^n / \Lambda_2^n$ can approach
any value as $n$ tends to infinity, i.e.,
\begin{equation}
R_2 = \frac{1}{n} \log | \mathcal{C}_2 | = \frac{1}{n} \log \left(
\frac{{\rm Vol} \left(\Lambda_2^{n}\right)}{{\rm
Vol}\left(\Lambda_C^{n}\right)} \right) = \gamma + o_n (1) \text,
\label{EQ:CodingRateRK}
\end{equation}
where $\mathcal{C}_2 = \{ \Lambda_C^n \md_2^n \}$ and $o_n(1)
\rightarrow 0$ as $n \rightarrow \infty$. Furthermore, the coding
rate of the nested lattice code associated with $\Lambda_C^n /
\Lambda_1^n$ is given by
\begin{equation*}
R_1 = \frac{1}{n} \log |\mathcal{C}_1| = \frac{1}{n} \log \left(
\frac{{\rm Vol} \left(\Lambda_1^{n} \right)}{{\rm Vol}
\left(\Lambda_C^{n} \right)} \right) = R_2 + \frac{1}{2} \log \left(
\frac{P_1}{P_2} \right) + o_n (1) \text,
\end{equation*}
where $\mathcal{C}_1 = \{ \Lambda_C^n \md_1^n \}$.
 \label{TH:LatticeChain}
\end{theorem}

\begin{proof}
See the proof of Theorem 2 in \cite{NamIT09Net}.
\end{proof}

\begin{figure} [t]
\begin{center}
\epsfig{file=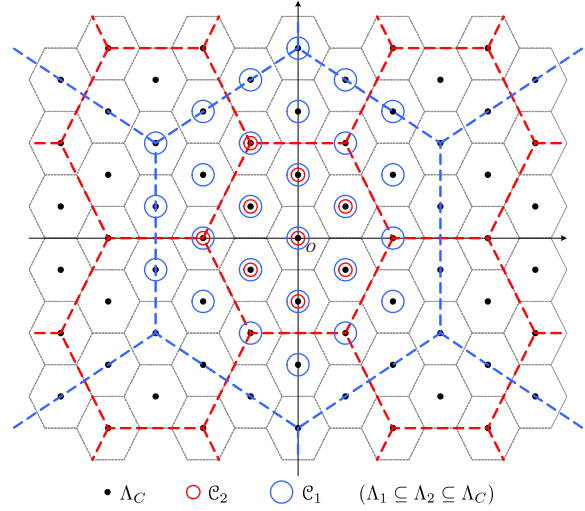, width=3in} \caption{Example of a lattice
chain and sets of coset leaders. $\mathcal{C}_2 \subseteq
\mathcal{C}_1 \subseteq \Lambda_C$.} \label{FIG:LatChain}
\end{center}
\end{figure}

For instance, a lattice chain and the corresponding sets of coset
leaders are visualized in Fig. \ref{FIG:LatChain} for the
two-dimensional case.

\psfull
\psdraft

\vspace{2mm}{\bf Encoding}\vspace{2mm}

Let us think of a lattice chain (more precisely, a sequence of
lattice chains) and sets of coset leaders as described in Theorem
\ref{TH:LatticeChain}. We use $\mathcal{C}_1$ and $\mathcal{C}_2$
for nodes 1 and 2 respectively. For node $i$, the message set
$\left\{1,\ldots,2^{nR_i} \right\}$ is one-to-one mapped to
$\mathcal{C}_i$. Thus, to transmit a message, node $i$ chooses ${\bf
W}_i \in \mathcal{C}_i$ associated with the message and sends
\begin{equation*}
{\bf X}_i = \left({\bf W}_i + {\bf U}_i \right) \md_i \text,
\end{equation*}
where ${\bf U}_i$ is a random dither vector with ${\bf U}_i \sim
{\rm Unif} (\mathcal{R}_i)$ and $\mathcal{R}_i$ denotes the Voronoi
region of $\Lambda_i$ (we suppressed the superscript `$^n$' for
simplicity). The dither vectors ${\bf U}_i$, $i \in \{1,2\}$, are
independent of each other and also independent of the messages and
the noise. We assume that each ${\bf U}_i$ is known to the source
nodes and the relay. Note that, due to the {\em crypto-lemma}
\cite{ForneyAllerton03}, ${\bf X}_i$ is uniformly distributed over
$\mathcal{R}_i$ and independent of ${\bf W}_i$. Thus, the average
transmit power of node $i$ is equal to $\sigma^2 (\Lambda_i)$, which
approaches $P_i$ as $n$ tends to infinity, and the power constraint
is met.

\vspace{2mm}{\bf Decoding}\vspace{2mm}

The received vector at the relay is given by
\begin{equation*}
{\bf Y}_R = {\bf X}_1 + {\bf X}_2 + {\bf Z}_R \text,
\end{equation*}
where ${\bf Z}_R = \left[ Z_R^{(1)}, \ldots, Z_R^{(n)} \right]^T$.
Upon receiving ${\bf Y}_R$, the relay computes
\begin{align*}
\tilde{\bf Y}_R &= \left(\alpha {\bf Y}_R - \sum_{j=1}^2 {\bf
U}_j\right) \md_1 \nonumber\\
&= \Bigg[ \sum_{j=1}^2 ({\bf W}_j + {\bf U}_j ) \md_j  -
\sum_{j=1}^2 {\bf X}_j \nonumber\\
& \;\;\;+ \alpha \sum_{j=1}^2 {\bf X}_j + \alpha {\bf Z}_R -
\sum_{j=1}^2
{\bf U}_j \Bigg] \md_1 \nonumber\\
&= \left( {\bf T} + \tilde{\bf Z}_R \right) \md_1 \text,
%\label{EQ:MLAN}
\end{align*}
where
\begin{align}
{\bf T} &= \left[ \sum_{j=1}^2 \left( {\bf W}_j - Q_j (
{\bf W}_j + {\bf U}_j) \right) \right] \md_1 \nonumber\\
&= \left[ {\bf W}_1 + {\bf W}_2 - Q_2 ( {\bf W}_2 + {\bf U}_2) \right] \md_1 \text, \label{EQ:EffCode}\\
\tilde{\bf Z}_R &= - (1- \alpha) ({\bf X}_1 + {\bf X}_2) + \alpha
{\bf Z}_R \text, \label{EQ:EffNoise}
\end{align}
$\alpha \in [0,1]$ is a scaling factor, and $Q_j (\cdot)$ denotes
the nearest neighbor lattice quantizer associated with $\Lambda_j$.
%i.e., $Q_j({\bf X}) = {\bf X} - {\bf X}\md_j$.
If we let $\alpha$ be the minimum mean-square error (MMSE)
coefficient
\begin{equation*}
\alpha = \frac{P_1 + P_2} {P_1 + P_2 + \sigma_R^2} \text,
\end{equation*}
the variance of the effective noise (\ref{EQ:EffNoise}) satisfies
\begin{equation*}
\frac{1}{n} E\left\{ \left\| \tilde{\bf Z}_R \right\|^2 \right\}
\leq \frac{(P_1 + P_2) \sigma_R^2} {P_1 + P_2 + \sigma_R^2} \text.
\label{EQ:NoiseVar}
\end{equation*}
From the chain relation of lattices in Theorem
\ref{TH:LatticeChain}, it follows that ${\bf T} \in \mathcal{C}_1$.
Moreover, using the crypto-lemma, it is obvious that ${\bf T}$ is
uniformly distributed over $\mathcal{C}_1$ and independent of
$\tilde{\bf Z}_R$ \cite[Lemma 2]{NamIT09Net}.

The relay attempts to recover ${\bf T}$ from $\tilde{\bf Y}_R$
instead of recovering ${\bf W}_1$ and ${\bf W}_2$ separately. Thus,
the lattice scheme inherits the idea of computation coding
\cite{NazerIT07} and physical-layer network coding
\cite{ZhangACM06}. Also, by not requiring the relay to decode both
messages, ${\bf W}_1$ and ${\bf W}_2$, we can avoid the multiplexing
loss \cite{KnoppIZS06} at the relay. The method of decoding we
consider is {\em Euclidean lattice decoding}
\cite{ErezIT04}-\cite{PoltyrevIT94}, which finds the closest point
to $\tilde{\bf Y}_R$ in $\Lambda_C$. Thus, the estimate of ${\bf T}$
is given by $\hat{\bf T} = Q_C \left( \tilde{\bf Y} \right)$, where
$Q_C(\cdot)$ denotes the nearest neighbor lattice quantizer
associated with $\Lambda_C$. Then, from the lattice symmetry and the
independence between ${\bf T}$ and $\tilde{\bf Z}_R$, the
probability of decoding error is given by
\begin{align}
p_e &= {\rm Pr} \left\{ \hat{\bf T} \neq {\bf T} %Q_C \left(\tilde{\bf Y}\right)
\right\} \nonumber\\
&= {\rm Pr} \left\{ \tilde{\bf Z}_R \md_1 \notin \mathcal{R}_C
\right\} \text, \label{EQ:ErrorProb}
\end{align}
where $\mathcal{R}_C$ denotes the Voronoi region of $\Lambda_C$. We
then have the following theorem.
\begin{theorem}
Let
\begin{equation*}
R_1^* = \left[ \frac{1}{2} \log \left( \frac{P_1}{P_1 + P_2} +
\frac{P_1}{\sigma_R^2} \right) \right]^+ \text.
\end{equation*}
For any $\bar{R}_1 < R_1^*$ and a lattice chain as described in
Theorem \ref{TH:LatticeChain} with $R_1$ approaching $\bar{R}_1$,
i.e., $R_1 = \bar{R}_1 + o_n(1)$, the error probability under
Euclidean lattice decoding (\ref{EQ:ErrorProb}) is bounded by
\begin{equation*}
p_e \leq e^{-n \left( E_P \left( 2^{2(R_1^* - \bar{R}_1)} \right) -
o_n(1) \right)} \text,
\end{equation*}
where $E_P(\cdot)$ is the Poltyrev exponent \cite{PoltyrevIT94}.

 \label{TH:LatticeAch}
\end{theorem}
%\begin{theorem}
%The decoding error probability $P_e$ in (\ref{EQ:ErrorProb})
%satisfies
%\begin{equation*}
%P_e \leq e^{-n \left( E_P \left( 2^{2(R_1^* - R_1)} \right) - o_n(1)
%\right)} \text,
%\end{equation*}
%where $E_P ( \cdot )$ is the Poltyrev exponent \cite{PoltyrevIT94},
%and
%\begin{equation*}
%R_1^* = \left[ \frac{1}{2} \log \left( \frac{P_1}{P_1 + P_2} +
%\frac{P_1}{\sigma_R^2} \right) \right]^+ \text.
%\end{equation*}
%\label{TH:LatticeAch}
%\end{theorem}

\begin{proof}
See the proof of Theorem 3 in \cite{NamIT09Net}.
\end{proof}

According to Theorem \ref{TH:LatticeAch}, the error probability
vanishes as $n \rightarrow \infty$ if $\bar{R}_1 < R_1^*$ since $E_p
(x) >0$ for $x>1$. This implies that the nested lattice code can
have any rate below $R_1^*$ for the reliable decoding of ${\bf T}$.
Thus, by c) of Theorem \ref{TH:LatticeChain} and Theorem
\ref{TH:LatticeAch}, the error probability at the relay vanishes as
$n \rightarrow \infty$ if %coding rate $R_i$, $i \in \{1,2\}$, can
%approach $R_i^*$ arbitrarily closely while keeping $P_e$ arbitrarily
%small for sufficiently large $n$, where
\begin{equation}
R_i < \left[ \frac{1}{2} \log \left( \frac{P_i}{P_1 + P_2} +
\frac{P_i}{\sigma_R^2} \right) \right]^+, \; i=1,2 \text.
\label{EQ:CodeRate}
\end{equation}
%This proves the uplink achievability.

\subsection{Downlink phase}

%Since nodes 1 and 2 know their own transmitted messages, the
%downlink channel of TRC falls into the class of broadcast channels
%with receiver side information \cite{OechteringIT08, XieCWIT07,
%WuISIT07}. The capacity of this class of channels can be achieved by
%the binning of messages. In our strategy, the binning is done
%structurally by the inherent group structure of lattice codes used
%in the uplink, which will be made clear at the end of this section. %The details of our downlink scheme are as follows.

We first generate $2^{nR_1}$ $n$-sequences with each element i.i.d.
according to $\mathcal{N}(0,P_R)$. These sequences form a codebook
$\mathcal{C}_R$. We assume one-to-one correspondence between each
${\bf t} \in \mathcal{C}_1$ and a codeword ${\bf X}_R \in
\mathcal{C}_R$. To make this correspondence explicit, we use the
notation ${\bf X}_R ({\bf t})$. After the relay decodes $\hat{\bf
T}$, it transmits ${\bf X}_R(\hat{\bf T})$ at the next block to
nodes 1 and 2. We now assume that there is no error in the uplink,
i.e., $\hat{\bf T}={\bf T}$. Under this condition, $\hat{\bf T}$ is
uniform over $\mathcal{C}_1$, and, thus, ${\bf X}_R(\hat{\bf T})$ is
also uniformly chosen from $\mathcal{C}_R$.

Upon receiving ${\bf Y}_1 = {\bf X}_R + {\bf Z}_1$, %node 1
%estimates the message of node 2 as $\hat{\bf W}_2 = {\bf w}$ if
%there exists a unique ${\bf w}$ in $\mathcal{C}_{R,1}$ such that
%$\left( {\bf X}_R ({\bf w}), {\bf Y}_1 \right)$ are jointly typical,
%where
%\begin{equation*}
%\mathcal{C}_{R,1} = \left\{ {\bf X}_R ({\bf t}
where ${\bf Z}_1 = \left[ Z_1^{(1)},\ldots,Z_1^{(n)} \right]^T$,
node 1 estimates the relay message $\hat{\bf T}$ as $\hat{\bf T}_1 =
{\bf t}_1$ if a unique codeword exists in $\mathcal{C}_{R,1}$ such
that $\left( {\bf X}_R ({\bf t}_1), {\bf Y}_1 \right)$ are jointly
typical, where
\begin{equation*}
\mathcal{C}_{R,1} = \left\{ {\bf X}_R ({\bf t}): {\bf t} = \left[
{\bf W}_1 + {\bf w}_2 - Q_2 ({\bf w}_2 + {\bf U}_2) \right] \md_1,
{\bf w}_2 \in \mathcal{C}_2 \right\} \text.
\end{equation*}
Then, from the knowledge of ${\bf W}_1$ and $\hat{\bf T}_1$, node 1
estimates the message of node 2 as
\begin{equation}
\hat{\bf W}_2 = \left( \hat{\bf T}_1 - {\bf W}_1 \right) \md_2
\text. \label{EQ:MsgEst1}
\end{equation}
Given $\hat{\bf T} = {\bf T}$, we have $\hat{\bf W}_2 = {\bf W}_2$
if and only if $\hat{\bf T}_1 = \hat{\bf T}$. Note that $\left|
\mathcal{C}_{R,1} \right| = 2^{nR_2}$. Thus, from the argument of
random coding and jointly typical decoding \cite{CoverText}, we have
\begin{equation}
\Pr \left\{ \hat{\bf T}_1 \neq \hat{\bf T} | \hat{\bf T} = {\bf T}
\right\} \rightarrow 0 \label{EQ:DLErr1}
\end{equation}
as $n \rightarrow \infty$ if
\begin{equation}
R_2 < \frac{1}{2} \log \left( 1 + \frac{P_R}{\sigma_1^2} \right)
\text. \label{EQ:DLRate1}
\end{equation}

Similarly, at node 2, the relay message is estimated to be $\hat{\bf
T}_2 = {\bf t}_2$ by finding a unique codeword in
$\mathcal{C}_{R,2}$ such that $\left( {\bf X}_R ({\bf t}_2), {\bf
Y}_2 \right)$ are jointly typical, where
\begin{equation*}
\mathcal{C}_{R,2} = \left\{ {\bf X}_R ({\bf t}): {\bf t} = \left[
{\bf w}_1 + {\bf W}_2 - Q_2 ({\bf W}_2 + {\bf U}_2) \right] \md_1,
{\bf w}_1 \in \mathcal{C}_1 \right\} \text.
\end{equation*}
Then the message of node 1 is estimated as
\begin{equation}
\hat{\bf W}_1 = \left[ \hat{\bf T}_2 - {\bf W}_2 + Q_2 ({\bf W}_2 +
{\bf U}_2 ) \right] \md_1 \text. \label{EQ:MsgEst2}
\end{equation}
Since $\left| \mathcal{C}_{R,1} \right| = 2^{nR_1}$, we have
\begin{equation}
\Pr \left\{ \hat{\bf T}_2 \neq \hat{\bf T} | \hat{\bf T} = {\bf T}
\right\} \rightarrow 0 \label{EQ:DLErr2}
\end{equation}
as $n \rightarrow \infty$ if
\begin{equation} R_1 < \frac{1}{2} \log
\left( 1+ \frac{P_R}{\sigma_2^2} \right) \text. \label{EQ:DLRate2}
\end{equation}
%Therefore, from (\ref{EQ:DLRate1}) and (\ref{EQ:DLRate2}), the
%downlink achievability is proved.

Note that, in the downlink, although the channel setting is
broadcast, nodes 1 and 2 achieve their point-to-point channel
capacities (\ref{EQ:DLRate1}) and (\ref{EQ:DLRate2}) without being
affected by each other. This is because of the side information on
the transmitted message at each node and the binning of message. In
our scheme, the relation in (\ref{EQ:EffCode}) represents how the
message pair $({\bf W}_1, {\bf W}_2)$ is binned to ${\bf T}$.

\subsection{Achievable rate region}

Clearly, the message estimates (\ref{EQ:MsgEst1}) and
(\ref{EQ:MsgEst2}) are exact if and only if $\hat{\bf T}_1 =
\hat{\bf T}_2 = {\bf T}$. Thus, the error probability is given by
\begin{align}
P_e &= \Pr \left\{\hat{\bf T_1} \neq {\bf T} \text{ or }
\hat{\bf T_2} \neq {\bf T} \right\} \nonumber\\
&\leq {\rm Pr} \left\{ \hat{\bf T}_1 \neq \hat{\bf T} \text{ or }
\hat{\bf T}_2 \neq \hat{\bf T} \text{ or } \hat{\bf T} \neq {\bf T}
\right\} \nonumber\\
&\leq \Pr \left\{\hat{\bf T} \neq {\bf T} \right\} + \Pr \left\{
\hat{\bf T}_1 \neq \hat{\bf T} | \hat{\bf T} = {\bf T} \right\} +
\Pr \left\{ \hat{\bf T}_2 \neq \hat{\bf T} | \hat{\bf T} = {\bf T}
\right\} \label{EQ:ErrProbTot}
\end{align}
By Theorem \ref{TH:LatticeAch}, the first term of
(\ref{EQ:ErrProbTot}) vanishes as $n \rightarrow \infty$ if $R_i <
R_i^*$, $i \in \{1,2\}$. Also, by (\ref{EQ:DLErr1}) and
(\ref{EQ:DLErr2}), the second and third terms also vanish as $n
\rightarrow \infty$ if (\ref{EQ:DLRate1}) and (\ref{EQ:DLRate2})
hold. Thus, the achievable rate region (\ref{EQ:AchRegion}) follows
from (\ref{EQ:CodeRate}), (\ref{EQ:DLRate1}), and
(\ref{EQ:DLRate2}).

\section{Conclusion} \label{SEC:Conclusion}

In this paper, we considered the Gaussian TRC. An achievable scheme
was presented based on nested lattice codes for the uplink and
structured binning for the downlink. The resulting achievable rate
region approaches to within $\frac{1}{2}$ bit of the cut-set bound
for all channel parameters, and the gap eventually vanishes in the
high
SNR regime. %In that sense, our scheme is at least as good as or
%better than the random coding schemes in \cite{AvestimehrAllerton08,
%GunduzAllerton08}. However, structured codes like the lattice codes
%are of practical interest because they reduce the complexity of
%encoding and decoding.
Though the capacity region is very nearly reached, the exact
capacity region of the Gaussian TRC is still an open problem. %Some kind of lattice coding with
%maximum-likelihood (ML) decoding at the relay may be a candidate of
%the capacity-achieving scheme \cite{Narayanan, NazerAllerton07}.
%However, with ML decoding, the analysis of error probability is not
%as easy as the Euclidean lattice decoding, and the extension is not
%straightforward.

% and it was conjectured that even the
%lattice coding with ML decoding may not be able to achieve the
%capacity region .

\end{document}